\documentclass{article}

\usepackage[preprint]{neurips_2025}

\usepackage[utf8]{inputenc} 
\usepackage[T1]{fontenc}    
\usepackage{hyperref}       
\usepackage{url}            
\usepackage{booktabs}       
\usepackage{amsfonts}       
\usepackage{nicefrac}       
\usepackage{microtype}      
\usepackage{xcolor}         
\usepackage{graphicx}
\usepackage{amsmath}  
\usepackage{adjustbox}
\usepackage{geometry}
\usepackage{rotating}
\usepackage{subcaption}
\usepackage{pifont}
\usepackage[inline]{enumitem}
\usepackage{array}
\usepackage{multirow}
\usepackage{tabularx}
\usepackage{bm}

\usepackage[utf8]{inputenc} 
\usepackage[T1]{fontenc}    
\usepackage{hyperref}       
\usepackage{url}            
\usepackage{booktabs}       
\usepackage{amsfonts}       
\usepackage{nicefrac}       
\usepackage{microtype}      
\usepackage{xcolor}         
\usepackage{graphicx}
\usepackage{amsmath}  
\usepackage{multirow} 
\usepackage{array} 
\usepackage[T1]{fontenc}
\usepackage{times}
\usepackage{tabularx}\title{UniMind: Unleashing the Power of LLMs for Unified Multi-Task Brain Decoding}
\usepackage{makecell}  
\usepackage{tcolorbox}
\usepackage{pifont}
\tcbuselibrary{breakable}

\author{%
\textbf{Weiheng Lu}$^{1,2}$\thanks{Equal contribution.}, 
\textbf{Zhouheng Yao}$^{1}$\footnotemark[1], 
\textbf{Jiamin Wu}$^{1,3}$\thanks{Correspondence.}, 
\textbf{Pengyu Zhu}$^{1}$\\
\textbf{Yuchen Zhou}$^{1}$,
\textbf{Weijian Mai}$^{1}$, 
\textbf{Qihao Zheng}$^{1}$, 
\textbf{Wanli Ouyang}$^{1,3}$,
\textbf{Chunfeng Song}$^{1}$ \\
$^1$Shanghai Artificial Intelligence Laboratory, 
$^2$Peking University, \\
$^3$The Chinese University of Hong Kong \\
}

\vspace{-10mm}

\begin{document}

\maketitle

\begin{abstract}
Decoding human brain activity from electroencephalography (EEG) signals is a central challenge at the intersection of neuroscience and artificial intelligence, enabling diverse applications in mental state assessment, clinical monitoring, and human–machine interaction. Recent efforts have extensively explored EEG-based brain foundation models for generalized brain decoding, employing large-scale training on multiple datasets. However, most of these attempts struggle with \textbf{generalizability} and fail to achieve satisfactory performance without \textbf{task-specific tuning} due to pronounced inherent heterogeneity among decoding tasks. 
To address these challenges, we present \textbf{\textit{UniMind}}, a general-purpose EEG foundation model for unified multi-task brain decoding by uniquely unleashing the power of LLMs to comprehend complex neural patterns.
UniMind enjoys several merits. First, we design a \textbf{Neuro-Language Connector} to bridge the modality gap between neural signals and LLMs, distilling and transforming the spatiotemporal neural patterns of EEG data into LLM-understandable representations.
Second, a \textbf{Task-aware Query Selection} module is proposed to inject task-awareness into the cross-modal alignment by dynamically generating task-adaptive query tokens, enabling the learning of task-relevant neural patterns across diverse tasks. Extensive experiments across 10 datasets demonstrate that UniMind substantially outperforms state-of-the-art multi-task decoding models (\textbf{11\%} gain on average), while also offering valuable neuroscientific insights into neural functional correlations across tasks. The code is available at \url{https://github.com/kaleidoyao/UniMind}.
\end{abstract}

\section{Introduction}

Electroencephalography (EEG) is a widely adopted technique used to measure and record electrical activity in the brain, where electrodes are placed on the scalp to detect and amplify the brain’s electrical signals. EEG plays a crucial role in Brain-Computer Interfaces (BCI) by providing a non-invasive, real-time measure of brain activity, and  also serves as a powerful tool for investigating brain's perceptual mechanisms. By analyzing neural patterns in EEG signals, many studies have been conducted to decode specific brain states, demonstrating strong potential across diverse applications such as seizure classification~\cite{intro_seizure}, sleep stage classification~\cite{intro_sleep}, motor imagery recognition~\cite{intro_motorimage}, abnormality detection~\cite{intro_abnormal}, emotion analysis~\cite{intro_emotion}, and acute
stress detection~\cite{intro_workload}.
\begin{figure}[t] 
\centerline{
  \includegraphics[width=1\linewidth]
  {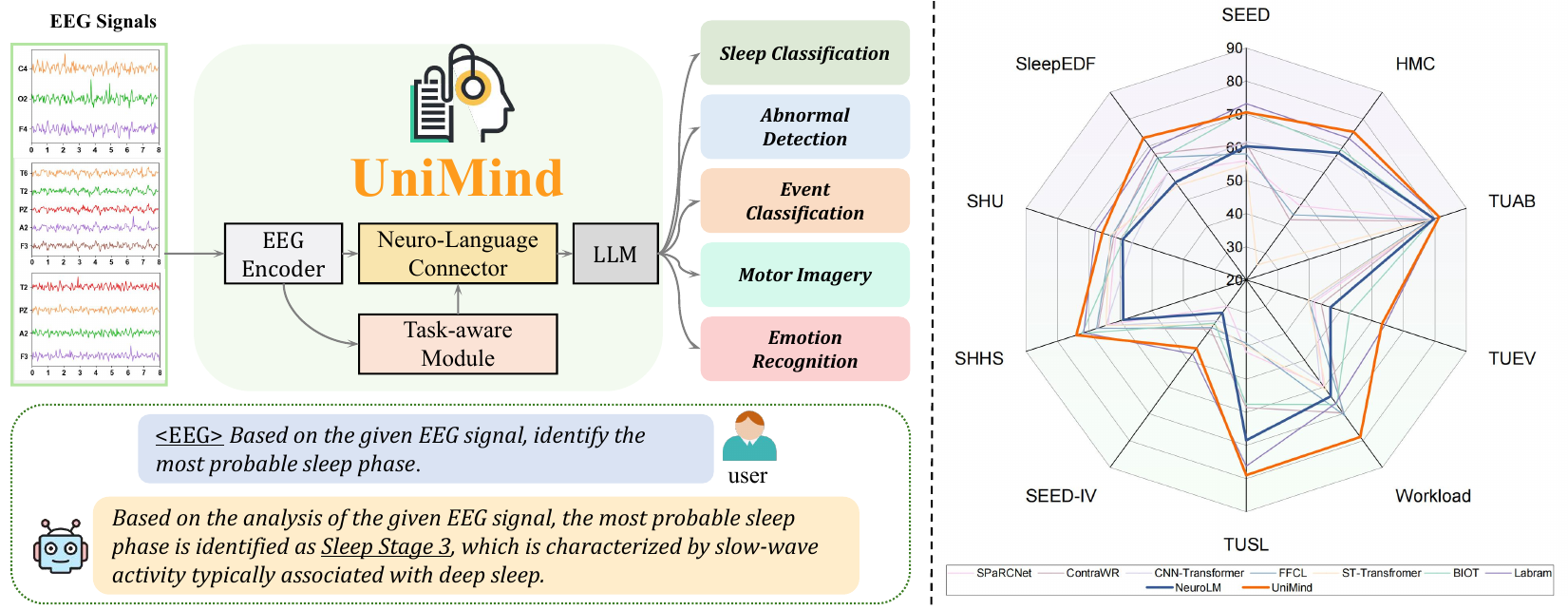}
  }
\caption{UniMind leverages LLMs to interpret brain signals, enabling multi-task EEG decoding without fine-tuning and demonstrating strong adaptability to task variability.}
\label{fig:teaser}
\vspace{-6mm}
\end{figure}

Despite the advances in EEG signal decoding, numerous deep learning models~\cite{jing2023SPaRCNet, 1d-cnnforeegsignals, dar2020cnnPhysiological, yang2022atd, kim2020ecg, jing2020automated, ObstructiveSleepApnoea, yang2023Electroencephalogram, CNNTransformer, li2022FFCL, song2021transformer, liu2022spatialtemporaltransformerseegemotion} have been confined to task-specific paradigms, due to the variations in EEG signal formats across different tasks. While effective for intended tasks, these models struggle to generalize to new tasks. Moreover, collecting task-specific EEG data is costly, making it impractical to build sufficient datasets for each task. These limitations highlight an urgent need for general-purpose EEG models that can learn transferable representations across  tasks. In response, foundation models such as BIOT~\cite{yang2023biot} and LaBraM~\cite{jiang2024Labram} have been proposed to acquire universal perceptual capabilities of EEG signals through large-scale self-supervised pre-training. However, these models require additional fine-tuning for each downstream task to outperform task-specific models, thus still struggling with generalizability and causing extra computational overhead.

To tackle the aforementioned challenges, recently proposed NeuroLM~\cite{jiang2025neurolm} developed a multi-task EEG foundation model that integrates Large Language Models (LLMs) using adversarial domain-level alignment and multi-channel autoregression to bridge EEG and language modalities.
While NeuroLM demonstrates the potential of LLMs in multi-task EEG understanding, its performance remains unsatisfactory compared to single-task methods, underperforming 
by more than 10\% on datasets such as SEED~\cite{zheng2015seed},  TUEV~\cite{harati2015tuab},  and SleepEDF~\cite{intro_sleep}.
To unlock LLMs' potential for decoding neural patterns and unifying diverse EEG tasks, two critical challenges need to be addressed:

\textbf{(1) Huge modality gap between neural signals and language.}
Unlike the rich, structured features of text tokens, EEG signals are distinguished by high noise,
sparse semantic information, and complex neural patterns that are difficult for LLMs to directly perceive. Alignment between EEG and language modalities is a prerequisite for LLMs to comprehend EEG signals and reason across modalities.
Existing methods use adversarial training to align EEG and text at the domain level, but the modality gap remains unresolved due to neglected semantic-level relations.
\emph{The core issue remains how to conduct effective \textbf{cross-modality bridging}  for alleviating the modality gap between EEG and LLMs.}

\textbf{(2) Heterogeneity across EEG decoding tasks.}
EEG tasks show significant heterogeneity~\cite{nonstationary_nature} due to varied configurations and cognitive mechanisms, resulting in diverse signal characteristics across datasets, such as electrode channels, trial durations, and spatiotemporal brain activity.
Given this, purely task-agnostic mixed-task tuning  may lead to degraded decoding performance on specific tasks.
Therefore, a task-aware mechanism is urgently required for generalizable brain decoding in the multi-task setting.
Moreover, task-agnostic methods overlook neural correlations across tasks, limiting insights into brain cognition.
\emph{How to develop a \textbf{task-aware mechanism} for achieving unified multi-task learning robust to EEG task heterogeneity without sacrificing individual task performance,} remains an open challenge. 

To address these challenges, we introduce \textbf{UniMind}, a general-purpose brain foundation model for unified multi-task brain decoding by unlocking the potential of LLMs in understanding brain patterns, as shown in Figure~\ref{fig:teaser}. To the best of our knowledge, it is the first multi-task EEG decoding model to match the performance of single-task approaches in one unified model.
To narrow the modality gap, we propose a \textbf{Neuro-Language Connector} (NLC) which acts as a compact, trainable bridge between the EEG encoder and the LLM, condensing essential brain patterns from sparse EEG data in a semantically meaningful way for the LLM to interpret. 
Specifically, to leverage the spatio-temporal nature of EEG signals~\cite{liu2022spatialtemporaltransformerseegemotion, spatial_temporal_featurefusion}, 
the neuro-language connector adopts a dual-branch architecture, with learnable query tokens that separately aggregate temporal dynamics and spatial dependencies from neural signals via cross-attention. The aggregated features are then aligned and mapped to the semantic space of a frozen LLM. By doing this, NLC transforms spatio-temporal neural patterns into interpretable features for LLMs, thereby enabling seamless neuro-language integration.
To facilitate effective multi-task learning across heterogeneous EEG tasks, we propose a \textbf{Task-aware Query Selection Module} (TQS)  to generate task-adaptive query tokens. TQS maintains spatial and temporal query pools containing multiple query tokens. A router mechanism is used to dynamically look up task-relevant queries from query pools based on input features. 
By allowing each task to adaptively choose its own queries, TQS promotes knowledge sharing and mutual enhancement across related tasks while mitigating interference from conflicting ones. 
Moreover, through the task selection mechanism, we uncover how diverse tasks are functionally organized across the brain by examining inter-task correlations, shedding light on how the brain regulates different cognitive functions.
Our contribution can be summarized as follows:
\begin{itemize}
  \item We propose \textbf{UniMind}, a general-purpose brain foundation model for unified multi-task brain decoding by integrating a spatio-temporal cross-modality bridging strategy between EEG and language, along with a task-aware mechanism.
  \item We propose a dual-branch neuro-language connector that encodes the temporal and spatial patterns of EEG signals into LLM-interpretable representations, along with a task-aware module that generates adaptive queries for task-relevant representations.
  \item Experiments demonstrate that UniMind outperforms the existing best  multi-task decoding model by \textbf{11\%} on average and is the first to achieve comparable or even superior performance to single-task decoding models across various tasks.
  \item Insights from a neuroscientific perspective are provided via query visualizations, revealing shared neural mechanisms and offering empirical support for knowledge sharing in multi-task brain decoding.
\end{itemize}

\section{Method}
\begin{figure}[t] 
\vspace{-2mm}
\centerline{
  \includegraphics[width=0.95\linewidth, height=83mm]
  {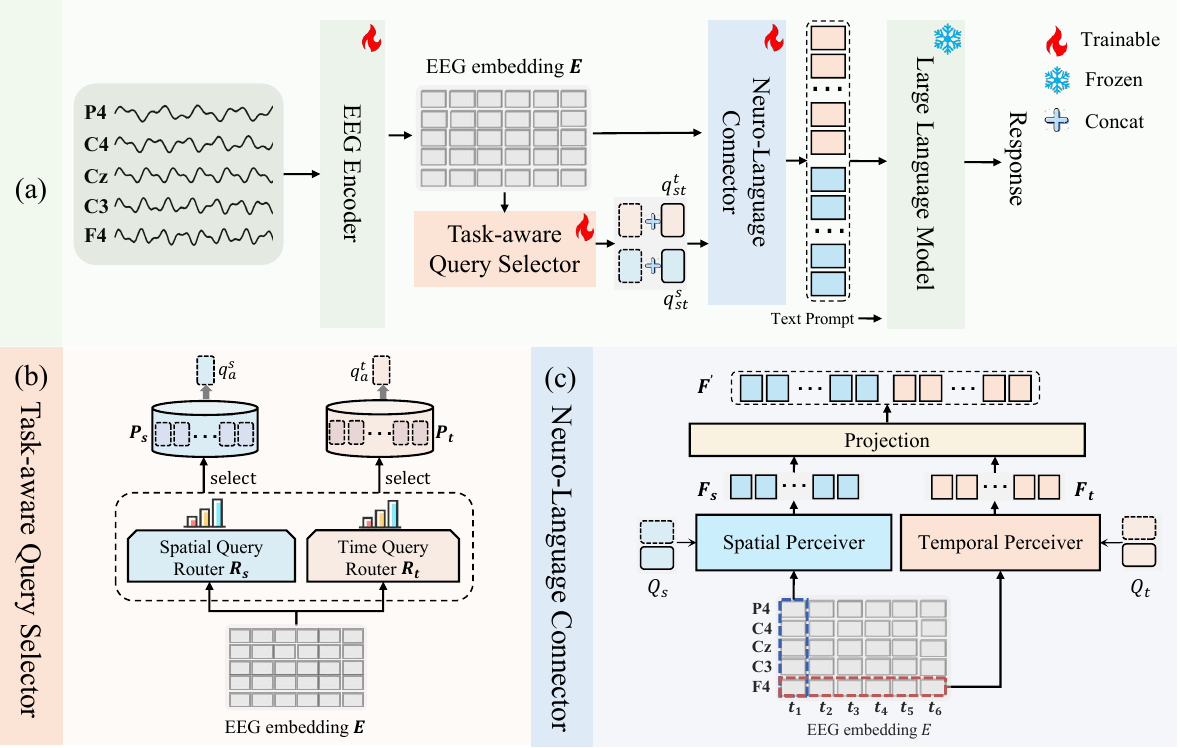}
}
\caption{Overview of the UniMind architecture. Raw EEG signals are encoded into EEG embeddings $\bm{E} $, which are processed by the Task-aware Query Selector to extract task-adaptive queries. These queries are combined with static queries and jointly processed with $\bm{E} $ by the Neuro-Language Connector, which aligns spatio-temporal neural features with the LLM’s semantic space. The resulting embeddings, together with task prompts, guide the LLM to generate text output.
}
\label{fig:architecture}
\vspace{-2mm}
\end{figure}

As illustrated in Figure~\ref{fig:architecture}, UniMind empowers large language models to understand brain signals from heterogeneous tasks by integrating two modules. (1) The \textbf{Neuro-Language Connector} (NLC)  (Section~\ref{sec:Neuro-Language Connector}) aims to bridge the modality gap between neural signals and language models by learning query tokens to interpret brain patterns.
(2) The \textbf{Task-aware Query Selection Module} (TQS) 
 (Section~\ref{sec:Task-aware Query Selection Module}) aims to enhance task adaptability by dynamically learning task-adaptive query tokens for feature selection across heterogeneous EEG tasks.

\subsection{Neuro-Language Connector}
\label{sec:Neuro-Language Connector}
Bridging the modality gap between EEG and language is the core issue in leveraging LLMs for multi-task brain decoding.
Unlike images and text, which have dense and well-structured data patterns, EEG signals differ significantly due to their sparse nature, low signal-to-noise ratio, and complex spatio-temporal neural patterns. 
EEG signals are characterized by temporal dynamics and multichannel structures~\cite{spatial_temporal_featurefusion, song2021transformer}.
To leverage these neural properties, the neuro-language connector is designed to condense and interpret spatiotemporal neural characteristics from noisy EEG signals, aligning them with the LLM's semantic space.
Next, we present the EEG encoder for extracting neural representations, followed by the design of our neuro-language connector.

\textbf{EEG Encoder.} 
Given an EEG signal sample \( X \in \mathbb{R}^{C \times S} \), where \( C \) and \( S \) respectively denote the number of electrode channels and temporal sampling points, a pre-trained EEG encoder is borrowed from LaBraM~\cite{jiang2024Labram}  to transform EEG signals with varying channels and lengths into standard token sequences, which consists of a temporal encoder for patch embedding and a Transformer encoder. Specifically, each EEG channel is divided into non-overlapping patches using a sliding window of length \( t \), resulting in \( T = \left\lfloor \frac{S}{t} \right\rfloor \) patches per channel. Finally, the signal is encoded as a sequence of patch token embeddings, with a prepended class token, resulting in \( \bm{E} \in \mathbb{R}^{(C \times T + 1) \times D_{\text{e}}} \).

\textbf{Neuro-Language Connector.} 
To bridge the gap between LLM and EEG signals with sparsity and low signal-to-noise ratio, 
the neuro-language connector aggregates  spatio-temporal features in a decoupled way via cross-attention.
Specifically, given the EEG embeddings $\bm{E} $ from the EEG encoder, the neuro-language connector designs two sets of learnable queries: temporal queries $Q_t \in \mathbb{R}^{n_{qt} \times D_{\text{e}}}$ and spatial queries $Q_s \in \mathbb{R}^{n_{qs} \times D_{\text{e}}}$, where $n_{qt}$ is the number of temporal queries and $n_{qs}$ is the number of spatial queries. These queries interact with EEG embeddings through separate cross-attention layers.
$Q_t$ captures temporal neural dynamics by attending to time-varying patterns within each channel, while $Q_s$ explores spatial  dependencies by analyzing spatial-wise channel activation patterns at each time step. 
We employ a cross-attention mechanism to discover and aggregate key brain patterns from noisy EEG signals, with the EEG token sequence serving as Keys and Values and learnable query tokens are Queries. Formally:
\begin{align}
\bm{F}_t &= \text{CrossAttn}(Q_t, \bm{E}_t) \in \mathbb{R}^{n_{qt} \times C \times D_{\text{e}}}, \quad
\bm{F}_s = \text{CrossAttn}(Q_s, \bm{E}_s) \in \mathbb{R}^{n_{qs} \times T \times D_{\text{e}}}.
\end{align}
Here, $\bm{E}_t \in \mathbb{R}^{C \times T \times D_{\text{e}}}$ and $\bm{E}_s \in \mathbb{R}^{T \times C \times D_{\text{e}}}$ are the reshaped EEG embeddings for temporal and spatial attention, respectively.
$\bm{F}_t$ and $\bm{F}_s$ are the temporally and spatially condensed EEG embeddings after attention-based interaction, which are then concatenated with the initial class token $\bm{F}_{\text{cls}} \in \mathbb{R}^{1 \times D_{\text{e}}}$, resulting in the neuro-semantic embeddings $\bm{F}$:
\begin{align}
\bm{F} = \text{Concat}(\bm{F}_t, \bm{F}_s, \bm{F}_{\text{cls}}) \in \mathbb{R}^{(n_{qt} \times C + n_{qs} \times T + 1) \times D_{\text{e}}}.
\end{align}

Finally, the neuro-semantic embeddings are projected to the embedding space of LLM to make them linguistically understandable to the LLM,
resulting in $\bm{F}'\in \mathbb{R}^{(n_{qt} \times C + n_{qs} \times T + 1) \times D_{\text{L}}}$,
where \( D_{\text{L}} \) denotes the LLM hidden size.
By transforming spatiotemporal neural patterns into semantically structured representations, neuro-language connector can enhance neuro-LLM alignment, empowering LLMs to interpret neural signals based to their underlying neurophysiological characteristics.

\subsection{Task-aware Query Selection Module}
\label{sec:Task-aware Query Selection Module}
Integrating multiple EEG tasks into one model is challenging due to heterogeneous signal characteristics across tasks.  
The shared architecture struggles to manage  task heterogeneity, leading to degraded performance on specific tasks and restricting mutual enhancement across related tasks in multi-task training.
To address this limitation, we introduce a Task-aware Query Selection Module (TQS) that integrates a task-aware mechanism to the cross-modality bridging process, dynamically learning query tokens for task-adaptive neural pattern mining and decoding.

To achieve dynamic query learning, TQS introduces a query pool $\bm{P}$ that contains a series of learnable query tokens $\bm{P}=[P_1, P_2, \cdots, P_{N_q}]$, and a query router $\bm{R}$ for query selection.  Specifically, TQS contains two branches for respective temporal and spatial query learning, denoted as $\bm{P}_t,\bm{R}_t$ and $\bm{P}_s,\bm{R}_s$.
The query router $\bm{R}$ generates routing scores $S$ over the query tokens in $\bm{P}$ conditioned on the input EEG embedding \( \bm{E} \).
The routing scores $S$ indicate how well each query fits to the input EEG features, allowing the model to generate task-adaptive queries tailored for the sample from the current task. 
Subsequently, the task-adaptive queries $q_a$ are selected from $\bm{P}$ based on routing scores:
\begin{align}
q_a^i = \text{TopK}(\bm{P}_i, S = R_i(\bm{E})), \quad i \in \{t, s\}, \quad q_a^i \in \mathbb{R}^{n_i \times D_e},
\end{align}
where $n_t$ and $n_s$ denote the number of task-adaptive queries for the temporal ($t$) and spatial ($s$) modalities, respectively. 
Notably, 
we incorporate an additional static query token $q_{st} \in \mathbb{R}^{1 \times D_e}$ besides $q_a$, which captures the stable characteristics of each task.
The final query token sequence can be formulated as: $Q_i = [q_a^i; q_{st}^i] \in \mathbb{R}^{(n_i + 1) \times D_e}, i \in \{t, s\}$. The resulting queries $Q_t$ and $Q_s$ are then fed into the NLC to extract task-relevant temporal and spatial features from EEG signals.  
TQS enables each task to select its own queries adaptively, fostering knowledge sharing and mutual enhancement across related tasks while reducing the interference risk from conflicting ones.

\subsection{Multi-Task Decoding with LLMs}
\label{sec:LLMdecoding}
\textbf{Multi-task instruction tuning dataset.}
We adopt instruction tuning for training the multi-task foundation model. To this end, we construct an instruction tuning dataset specifically tailored for brain decoding, 
comprising five diverse task types and containing a total of 929k samples.
To further enrich the instruction diversity, we design ten distinct prompts for each task type. For every sample, one of these prompts is randomly selected and paired as a task-specific instruction. Further details regarding the instruction templates are provided in the \textbf{Appendices}.

\textbf{LLM Training.} 
We formulate the EEG decoding process as a conditional language modeling task using the instruction tuning dataset, where the model is expected to generate the correct label or description conditioned on EEG signals and the instruction. The LLM input sequence is formulated as the concatenation of neural-semantic representation \( F' \), a special separator token \( \text{[SEP]} \) and the instruction prompt \( P \) describing the task:
$X = [F', \text{[SEP]}, P]$.
The model is trained to generate the ground truth label \( Y \) by predicting the corresponding target token sequence \( t_a \), and is optimized using a causal cross-entropy loss for LLM training:
\begin{equation}
\mathcal{L} = - \sum_{i=1}^{L} \log p(t_{a,i} \mid X, t_{a,<i}).
\end{equation}
Where $t_{a,i}$ is the predicted token and $t_{a,<i}$ are the preceding tokens.

\section{Experiments}
\subsection{Experimental Setup}

\textbf{Datasets \& Metrics.} 
To comprehensively evaluate \textit{UniMind}, we adopt ten publicly available EEG datasets across five task domains: sleep stage classification (HMC~\cite{alvarez2021hmc}, SleepEDF~\cite{kemp2000sleepedf}, SHHS~\cite{quan1997shhs}), emotion recognition (SEED~\cite{zheng2015seed}, SEED-IV~\cite{zheng2019seediv}), clinical anomaly detection (TUAB~\cite{harati2015tuab}, TUEV~\cite{harati2015tuab}, TUSL~\cite{von2017tusl}), cognitive workload classification (Workload~\cite{zyma2019electroencephalograms}), and motor imagery classification (SHU~\cite{ma2022shu}). All experiments follow the training, validation, and test splits to ensure fair and consistent evaluation. Detailed information for each dataset, additional preprocessing, and data split strategies are provided in the \textbf{Appendices}. Given the class imbalance in EEG datasets, we evaluate model performance using three primary metrics: \textbf{Balanced Accuracy}, which averages recall across classes and is robust to imbalance; \textbf{Cohen’s Kappa}, which measures agreement between predicted and true labels while accounting for chance; and \textbf{Weighted F1}, which balances precision and recall, weighted by class frequency.

\textbf{Implementation Details.} We utilize the EEG encoder from LaBraM~\cite{jiang2024Labram}  and adopt InternLM2.5-7B~\cite{cai2024internlm2} as the language model. 
In NLC, each branch employs an 8-head cross-attention layer, followed by a two-layer MLP.
The query router $\bm{R}$ is implemented as a two-layer MLP, and the query pool size is set to 16. Static queries use 1 token for both spatial and temporal dimensions, while task-adaptive queries use 2 spatial tokens and 1 temporal token.
We discuss the choice of pool size and the number of queries in detail in ~\ref{sec: ablation}. During training, the parameters of the LLM are frozen to preserve its pre-trained language understanding, while the other components are trainable. 
All experiments are conducted on a machine with eight NVIDIA A800 GPUs (80GB). Additional training details and computational resource analysis are provided in the \textbf{Appendices}.

\textbf{Compared Methods.} To ensure a comprehensive evaluation, we compare \textit{UniMind} against both non-foundation and foundation model baselines. \textbf{(1)~Non-foundation models.} We include state-of-the-art deep learning models such as SPaRCNet~\cite{jing2023SPaRCNet}, ContraWR~\cite{yang2023Electroencephalogram}, CNN-Transformer (CNN-Trans)~\cite{CNNTransformer}, FFCL~\cite{li2022FFCL}, and ST-Transformer (ST-Trans)~\cite{song2021transformer}. These models typically adopt supervised learning strategies that are specifically designed for individual tasks or datasets. \textbf{(2)~Foundation models.} We consider NeuroLM~\cite{jiang2025neurolm}, BIOT~\cite{yang2023biot}, and LaBraM~\cite{jiang2024Labram}. NeuroLM is available in three parameter scales (B, L, XL), and we report the best-performing version for each dataset.  For fair comparison, we follow NeuroLM's data splits and evaluation metrics on SEED, HMC, TUEV, and TUSL, and directly report their published results as provided in NeuroLM. For the remaining datasets, we re-implement the baseline models using our own data splits and evaluation metrics.

\begin{table}[t]
\centering
\caption{Performance comparison of various models across different EEG datasets. We \underline{underline} the best performance among multi-task models and \textbf{bold} the best overall.}
\renewcommand{\arraystretch}{0.85}
\label{tab:main_exp}
\setlength{\tabcolsep}{2pt}
\scriptsize
\resizebox{1.0\textwidth}{!}{
\begin{tabular}
{ll|p{35pt}<{\centering}p{30pt}<{\centering}p{35pt}<{\centering}p{28pt}<{\centering}p{35pt}<{\centering}p{28pt}<{\centering}p{30pt}<{\centering}|p{32pt}<{\centering}p{32pt}<{\centering}}
\toprule
\multicolumn{2}{c|}{} & \multicolumn{7}{c}{\textbf{Single-task}} & \multicolumn{2}{|c}{\textbf{Multi-task}} \\
\cmidrule(lr){3-9} \cmidrule(lr){10-11}
\textbf{Dataset} & \textbf{Metrics} & \textbf{SPaRCNet} & \textbf{ContraWR} & \textbf{CNN-Trans} & \textbf{FFCL} & \textbf{ST-Trans} & \textbf{BIOT} & \textbf{LaBraM} & \textbf{NeuroLM} & \textbf{UniMind} \\
\midrule
\multirow{2}{*}{SEED} & B-Acc & 55.96 & 61.06 & 61.61 & 58.08 & 54.79 & 70.97 & \textbf{73.18} & 60.34 & \underline{70.55} \\
& F1-W  & 55.85 & 61.37 & 61.50 & 57.43 & 55.05 & 71.34 & \textbf{73.54} & 60.63 & \underline{70.98} \\
\midrule
\multirow{2}{*}{HMC} & B-Acc & 47.56 & 42.42 & 65.73 & 44.27 & 25.59 & 68.62 & 72.86 & 67.37 & \textbf{\underline{75.27}} \\
& F1-W  & 41.08 & 29.87 & 68.96 & 29.02 & 14.28 & 70.91 & 75.54 & 71.26 & \textbf{\underline{77.40}} \\
\midrule
\multirow{2}{*}{TUAB} & B-Acc & 78.69 & 80.17 & 79.53 & 78.19 & 79.66 & 79.59 & 81.40 & 79.69 & \textbf{\underline{81.76}} \\
& F1-W  & 75.13 & 80.65 & 78.76 & 77.83 & 80.90 & 78.82 & 81.47 & 78.93 & \textbf{\underline{82.03}} \\
\midrule
\multirow{2}{*}{TUEV} & B-Acc & 41.61 & 43.84 & 40.87 & 39.79 & 39.84 & 52.81 & \textbf{64.09} & 46.79 & \underline{63.19} \\
& F1-W  & 70.24 & 68.93 & 68.54 & 67.83 & 68.23 & 74.92 & \textbf{83.12} & 73.59 & \underline{78.04} \\
\midrule
\multirow{2}{*}{Workload} & B-Acc & 59.77 & 69.66 & 57.93 & 70.69 & 61.03 & 66.55 & 66.09 & 63.45 & \textbf{\underline{78.67}} \\
& F1-W  & 54.60 & 69.33 & 58.63 & 72.54 & 60.67 & 51.66 & 64.72 & 66.57 & \textbf{\underline{78.65}} \\
\midrule
\multirow{2}{*}{TUSL} & B-Acc & 41.85 & 58.57 & 35.75 & 39.19 & 40.00 & 57.58 & 76.25 & 68.45 & \textbf{\underline{78.95}} \\
& F1-W  & 35.00 & 54.58 & 22.35 & 21.20 & 37.93 & 23.94 & \textbf{76.14} & 68.39 & \underline{75.40} \\
\midrule
\multirow{2}{*}{SEED-IV} & B-Acc & 29.88 & 38.38 & 35.21 & 37.81 & 36.93 & 36.19 & \textbf{47.63} & 32.30 & \underline{45.56} \\
& F1-W  & 32.05 & 40.21 & 36.57 & 39.76 & 36.95 & 42.76 & \textbf{49.14} & 34.65 & \underline{43.58} \\
\midrule
\multirow{2}{*}{SleepEDF} & B-Acc & 60.16 & 67.05 & 60.29 & 65.66 & 55.17 & 64.95 & 68.96 & 56.40 & \textbf{\underline{72.98}} \\
& F1-W  & 58.61 & 66.92 & 58.96 & 64.79 & 53.18 & 60.91 & 87.30 & 54.02 & \textbf{\underline{88.23}} \\
\midrule
\multirow{2}{*}{SHHS} & B-Acc & 63.93 & 67.01 & 64.63 & 67.59 & 64.63 & 72.22 & 71.69 & 59.15 & \textbf{\underline{74.00}} \\
& F1-W  & 61.82 & 56.80 & 63.78 & 67.07 & 63.30 & 83.96 & 82.90 & 63.54 & \textbf{\underline{84.20}} \\
\midrule
\multirow{2}{*}{SHU} & B-Acc & 62.15 & 62.13 & 56.25 & 62.82 & 63.39 & 59.16 & \textbf{67.90} & 59.36 & \underline{65.77} \\
& F1-W  & 62.15 & 57.51 & 55.86 & 62.78 & 63.25 & 55.51 & \textbf{67.84} & 56.03 & \underline{65.73} \\
\midrule
& \textbf{Average} & 54.16 & 59.03 & 55.78 & 56.41 & 52.10 & 62.86 & 69.01 & 59.33 & \textbf{\underline{70.67}} \\
\bottomrule
\end{tabular}
}
\vspace{-6mm}
\end{table}

\subsection{Comparison with SOTA Methods}

Table~\ref{tab:main_exp} compares UniMind with state-of-the-art methods on ten brain decoding datasets. UniMind achieves the highest average performance, leading multi-task brain decoding. 
Next we provide a comparative analysis between UniMind and both multi-task and single-task models. We present the results for Balanced Accuracy and Weighted F1 here, while the complete results, including full metrics and error analysis, are provided in the \textbf{Appendices}.

\textbf{Comparison with multi-task models.} We first compare our method with \textit{NeuroLM} in the same multi-task setting. \textit{UniMind} consistently outperforms \textit{NeuroLM} across all datasets by a large margin, achieving an average performance gain of \textbf{11\%}. 
This improvement stems from UniMind's deep semantic alignment through the neuro-language connector, which enhances the ability of the LLM to interpret neural signals for multi-task decoding, unlocking its neural inference ability.  Besides, the task-aware query selection module boosts the model's task adaptability to heterogeneous decoding tasks.
Notably, our method achieves remarkable performance 
on the challenging SEED-IV and TUEV datasets, substantially improving their initially low balanced accuracies (\textit{i.e.,} \textbf{$\uparrow$ 13.26\%} and \textbf{$\uparrow$16.4\%}). Another interesting phenomenon is that \textit{UniMind} demonstrates a substantial improvement of \textbf{15.22\%} in balanced accuracy on the small-scale Workload dataset with only 1k samples. 
These improvements stem from TQS's task-adaptive query selection from a shared pool, which promotes knowledge transfer and mutual enhancement among tasks. This mechanism facilitates effective multi-task training, boosting performance on both complex and data-limited tasks.

\textbf{Comparison with single-task models.} We further compare UniMind with several state-of-the-art single-task models. It should be noted that multi-task learning is inherently more challenging. Therefore, this comparison is not entirely fair and is intended primarily as a reference for performance evaluation.  \textit{UniMind} achieves the best balanced accuracy among all single-task models on HMC, TUAB, SleepEDF, SHHS, Workload, and TUSL, and also obtains the \textbf{best} average score across all datasets. On the remaining datasets SEED, SEED-IV, TUEV, and SHU, it ranks second only to LaBraM, with a marginal performance gap. These results demonstrate that \textit{UniMind} is capable of matching or even surpassing leading single-task baselines in a unified model. The effective brain-language alignment reduces the performance gap compared to task-specific tuning. Besides, the knowledge sharing facilitated by the task-aware query selector allows each dataset to benefit from semantically similar datasets, a capability that single-task models do not possess.

\subsection{Ablation Study}
\label{sec: ablation}

\begin{table}[t]
\centering
\vspace{-2mm}
\caption{Ablation of Model Components (Balanced Accuracy).}
\label{table:ablation component}
\begin{adjustbox}{max width=\textwidth}
\begin{tabular}{l|ccccccccccc}
\toprule
\textbf{Model Variant} & \textbf{AVG} &\textbf{SEED} & \textbf{HMC} & \textbf{TUAB} & \textbf{TUEV} & \textbf{Workload} & \textbf{TUSL} & \textbf{SEED-IV} & \textbf{SleepEDF} & \textbf{SHHS} & \textbf{SHU} \\
\midrule
Baseline & 65.77 & 64.35 & 73.87 & 79.63 & 55.49 & 65.33 & 72.65 & 42.74 & 71.07 & 68.59 & 63.96 \\
+NLC & 68.38 & 69.66 & 74.43 & 80.97 & 59.79 & 70.66 & 76.69 & 43.40 & 72.01 & 71.18 & 64.90 \\
+NLC+TQS  & \textbf{70.67} & \textbf{70.55} & \textbf{75.27} & \textbf{81.76} & \textbf{63.19} & \textbf{78.67} & \textbf{78.95} & \textbf{45.56} & \textbf{72.98} & \textbf{74.00} & \textbf{65.77} \\
\bottomrule
\end{tabular}
\end{adjustbox}
\vspace{-5mm}
\end{table}
\begin{table}[t]
\centering
\caption{Ablation of Using Static and Task-adaptive Queries (Balanced Accuracy).}
\label{table:ablation_querytype}
\begin{adjustbox}{max width=\textwidth}
\begin{tabular}{l|ccccccccccc}
\toprule
\textbf{Query Type} & \textbf{AVG} & \textbf{SEED} & \textbf{HMC} & \textbf{TUAB} & \textbf{TUEV} & \textbf{Workload} & \textbf{TUSL} & \textbf{SEED-IV} & \textbf{SleepEDF} & \textbf{SHHS} & \textbf{SHU} \\
\midrule
Static & 68.37 & 69.66 & 74.43 & 80.97 & 59.79 & 70.66 & 76.69 & 43.40 & 72.01 & 71.18 & 64.90 \\
Adaptive & 69.34  & 68.51 & 74.62 & 80.62 & 62.83 & 74.00 & 77.07 & 44.77 & 72.26 & 73.39 & 65.30 \\
Static+Adaptive & \textbf{70.67} & \textbf{70.55} & \textbf{75.27} & \textbf{81.76} & \textbf{63.19} & \textbf{78.67} & \textbf{78.95} & \textbf{45.56} & \textbf{72.98} & \textbf{74.00} & \textbf{65.77}  \\
\bottomrule
\end{tabular}
\end{adjustbox}
\vspace{-3mm}
\end{table}
\begin{table}[t]
\vspace{-3mm}
\centering
\caption{The Impact of Number of Task-adaptive Queries (Balanced Accuracy).}
\label{table:ablation query counts}
\begin{adjustbox}{max width=0.95\textwidth}
\begin{tabular}{cccccccccccccc}
\toprule
\textbf{$n_s$} & \textbf{$n_t$} & \textbf{AVG} & \textbf{SEED} & \textbf{HMC} & \textbf{TUAB} & \textbf{TUEV} & \textbf{Workload} & \textbf{TUSL} & \textbf{SEED-IV} & \textbf{SleepEDF} & \textbf{SHHS} & \textbf{SHU} \\
\midrule
1 & 2 & 66.79 & 66.04 & 74.71 & 79.62 & 50.15 & 76.00 & 74.77 & 43.64 & 69.87 & 69.37 & 63.76 \\
1 & 2 & 68.44 & 68.30 & 74.00 & 80.77 & 55.41 & \textbf{78.67} & \textbf{77.95} & 44.05 & 70.51 & 70.64 & 64.14 \\
2 & 1 & \textbf{69.34} & 68.51 & 74.62 & 80.62 & \textbf{62.83} & 74.00 & 77.07 & \textbf{44.77} & 72.26 & \textbf{73.39} & \textbf{65.30} \\
2 & 2 & 68.39 & 68.23 & \textbf{74.67} & 80.34 & 57.89 & 73.18 & 77.42 & 44.36 & 72.07 & 72.50 & 63.27 \\
2 & 4 & 67.24 & 67.41 & 74.06 & 80.46 & 59.27 & 74.66 & 65.86 & 43.50 & \textbf{72.98} & 71.69 & 62.47 \\
4 & 2 & 67.75 & 67.44 & 74.11 & 80.67 & 56.70 & 71.33 & 77.10 & 44.51 & 71.39 & 71.52 & 62.76 \\
4 & 4 & 68.14 & \textbf{68.55} & 74.10 & \textbf{81.59} & 57.38 & 74.67 & 75.02 & 44.10 & 71.62 & 71.58 & 62.80 \\
\bottomrule
\end{tabular}
\end{adjustbox}
\vspace{-3mm}
\end{table}

\textbf{Analysis of Key Model Components.}
We conduct an ablation study to evaluate the contributions of key model components, as shown in Table~\ref{table:ablation component}. The baseline model employs a fixed number of learnable queries to extract features from the entire EEG sequence. Introducing the NLC module results in substantial performance gains across all datasets, with improvements around 5\% on more challenging tasks such as SEED and TUEV. This is due to the NLC’s ability to capture temporal and spatial patterns in EEG embeddings and convert them into semantic representations that improve neuro-LLM alignment.
The addition of the TQS further enhances performance, yielding gains of 3.40\% on TUEV, 2.82\% on SHHS, and 8.01\% on Workload. These improvements stem from the task-aware mechanism that enables adaptive query learning across tasks, promoting knowledge sharing among related tasks while reducing inter-task interference in cross-modality bridging.

\textbf{Comparison of Different Types of Queries.} We compare different query types used in the NLC module to evaluate the impact of static and task-adaptive query selection, as shown in Table~\ref{table:ablation_querytype}. 
Compared to using only static queries, employing task-adaptive queries selected by the TQS module yields notable improvements on most datasets, with gains of 3.04\% on TUEV and 3.34\% on Workload. This improvement is primarily due to the ability of each task to adaptively select queries that better capture task-relevant temporal and spatial features.
Furthermore, combining static and task-adaptive queries yields the best overall performance across all datasets.

\textbf{Analysis of the Number of Task-adaptive Queries.}  We analyze the impact of the number of dynamic queries in Table~\ref{table:ablation query counts}, where $n_s$ and $n_t$ denote the numbers of task-adaptive queries $q_a^s$ and $q_a^t$. The results show that increasing the number of queries does not always improve performance. Too few queries limit the model’s capacity to capture diverse EEG patterns, while too many queries may introduce noise because some queries may not align well with the EEG embeddings.
While the optimal setup varies by dataset, the best overall results occur with \( n_s = 2 \) and \( n_t = 1 \).

\begin{figure}[t]
  \centering
    \includegraphics[width=1.0\textwidth]{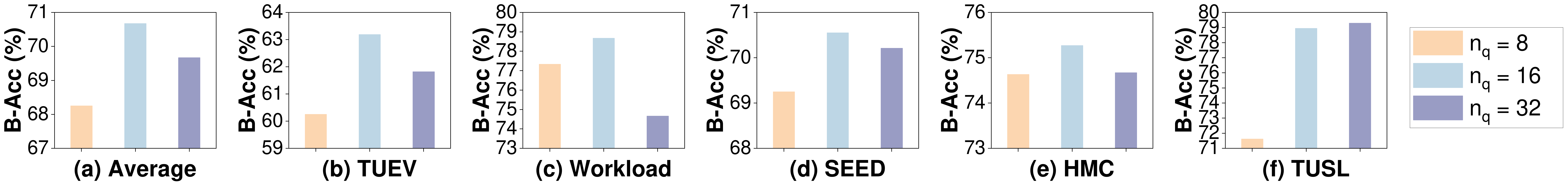}
       \vspace{-5mm}
  \caption{Configuration of Query Pool Size $n_q$.}
  \label{fig: poolsize}
   \vspace{-5mm}
\end{figure}

\textbf{Analysis of Query Pool Size.} 
We study the effect of the query pool sizes $\bm{P}_t$ and $\bm{P}_s$ (of equal size, denoted as $n_q$) in Figure~\ref{fig: poolsize}.
When the pool is too small, task-specific selections tend to overlap, which can lead to interference between tasks. Conversely, an overly large pool reduces query overlap among similar tasks, weakening the mutual enhancement that could occur between related tasks. Setting the pool size to 16 yields the best overall performance by balancing diversity and shared representation.

\begin{figure}[t]
  \centering
    \includegraphics[width=1.0\textwidth]{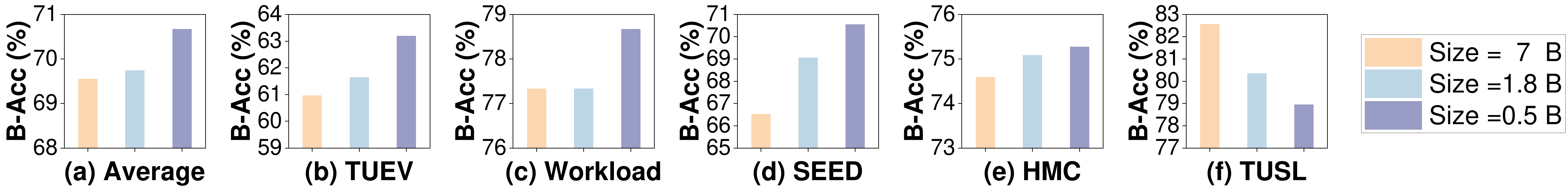}
 \vspace{-4mm}
  \caption{Ablation Study on Model Parameter Size.}
  \label{fig:parametersize}
  \vspace{-6mm}
\end{figure}


\textbf{Analysis of Model Parameter Size.} Figure~\ref{fig:parametersize} investigates the impact of different model sizes on performance. We compare InternLM models with approximately 0.5B, 1.8B, and 7B parameters. We find that models with more parameters tend to bring consistent improvements on more challenging datasets such as SEED and TUEV, likely due to their stronger capacity to model complex patterns. In contrast, the performance of TUSL degrades as model size increases. This may be attributed to the limited amount of training data (245 samples), which makes larger models more prone to overfitting.
\subsection{Task-level Query Analysis}
In this section, we explore task-level query selection patterns and inter-task correlations
by query visualization, shedding light on underlying neural functional mechanisms. We further investigate the mutual enhancement effect, examining whether functionally similar tasks benefit from joint training.

\begin{figure}[t] 
\vspace{-2mm}
\centerline{
  \includegraphics[width=1\linewidth]
  {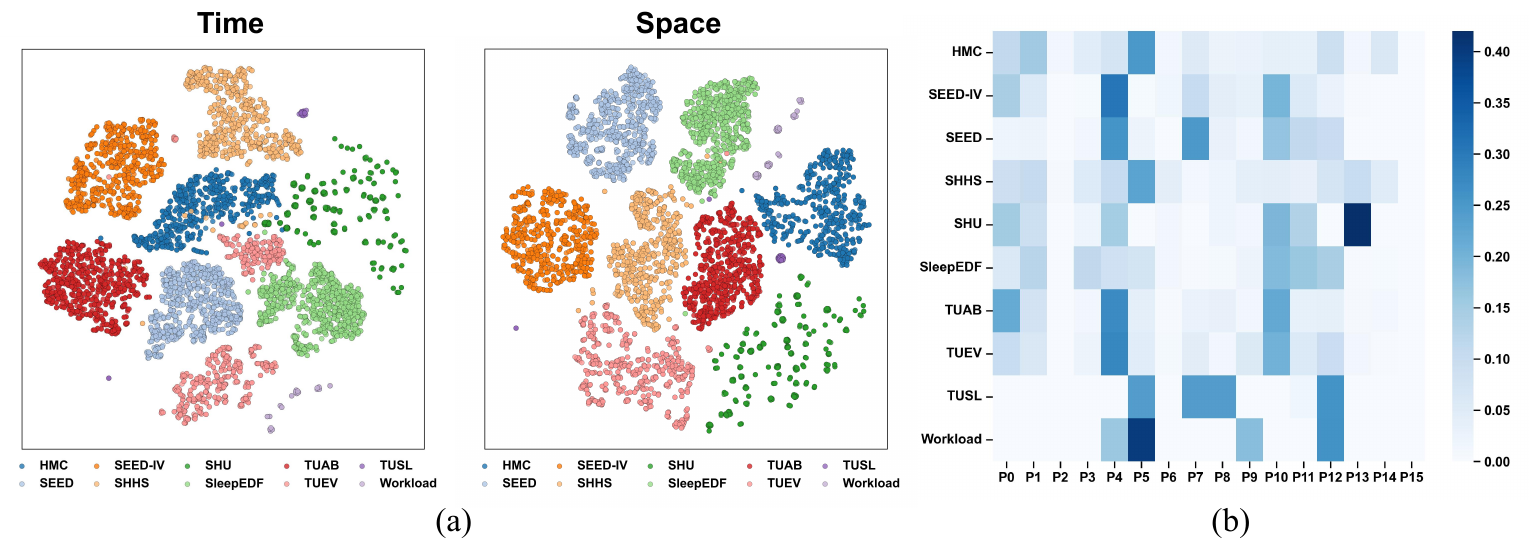}
\vspace{-2mm}
}
\caption{(a): t-SNE-based visualization of neural routing distributions across datasets; (b): task-adaptive spatial query distributions across datasets.
}
\label{fig:tsne}
\vspace{-3mm}
\end{figure}

\textbf{Visualization of Task-adaptive Query Selection.} To investigate the working mechanism of the Task-adaptive Query Selection module,  we visualize the spatial and channel routing scores \( S \) from query routers \( \bm{R} \) across EEG datasets using t-SNE. The query scores reflect the query selection pattern in generating task-adaptive queries.
As shown in Figure~\ref{fig:tsne}~(a), the neural routing distribution presents a clear \textbf{task-aware pattern} for both temporal and spatial routers, with samples from different datasets forming clearly separated clusters. The well-separated distribution suggests that TQS is capable of dynamically selecting queries for task-adaptive neuro-LLM alignment across heterogeneous tasks.
 
Furthermore, we analyze the selection frequencies of task-adaptive spatial queries across datasets to reveal each task's preferences, as illustrated in Figure~\ref{fig:tsne}~(b). The results indicate that query selection exhibits clear task-dependent trends. For example, emotion recognition and clinical event detection tasks (SEED, SEED-IV, TUAB, TUEV) tend to select queries \( P_4 \) and \( P_{10} \), while sleep stage classification tasks (SHHS, HMC) consistently favor \( P_5 \).  These findings suggest that the queries in  $\bm{P}$ effectively capture diverse underlying neural characteristics across brain regions.

\begin{figure}[t] 
\centerline{
  \includegraphics[width=1.0\linewidth]
  {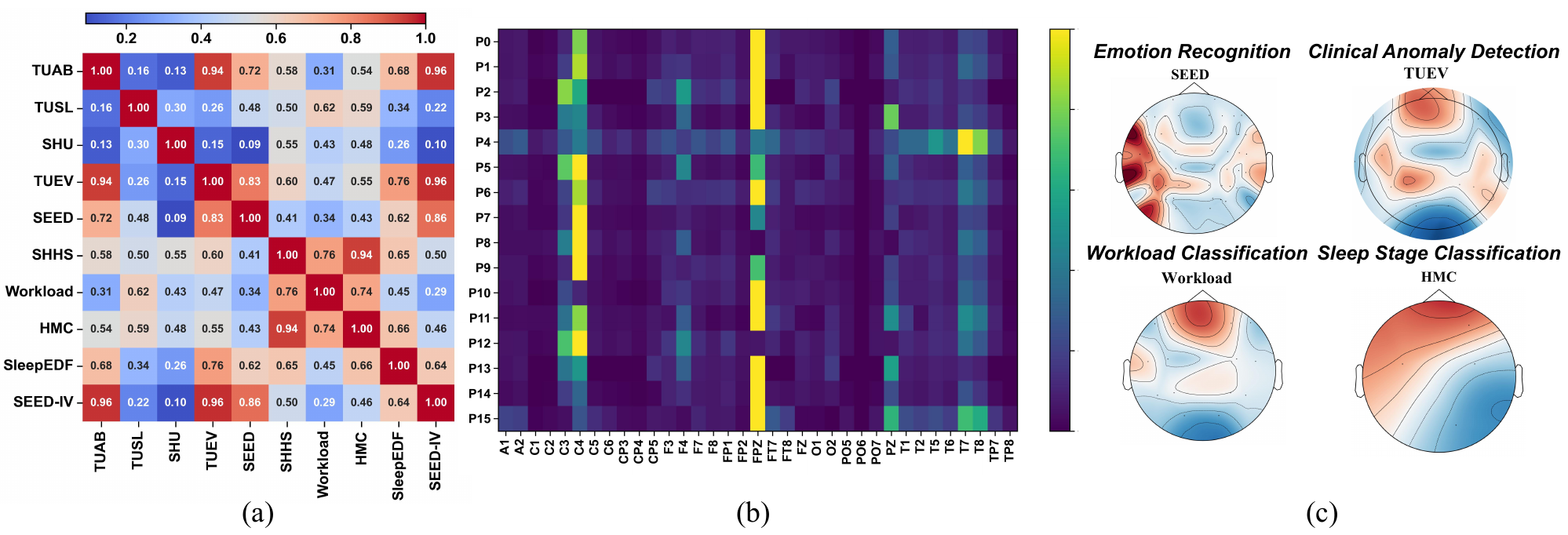}
}
\vspace{-2mm}
\caption{(a): Similarity of task-adaptive query distributions across tasks; (b): The attention values of task-adaptive queries across different channels; (c): Topography visualization on different tasks. 
}
\label{fig:Task-aware Query Selection Module}
\vspace{-5mm}
\end{figure}

\textbf{Neural Correlations across Tasks.} 
To capture underlying neural correlations across tasks, we measure similarities in query selection frequencies among EEG tasks to reflect neural patterns. Here we analyze the spatial query tokens.
As illustrated in Figure~\ref{fig:Task-aware Query Selection Module}, we highlight two key observations. (1) Datasets/tasks belonging to the same task domain
tend to exhibit much higher similarity scores. For example, TUAB and TUEV (clinical), SEED and SEED-IV (emotion), as well as SHHS and HMC (sleep), show remarkably consistent selection patterns. These findings indicate that TQS can help capture consistent neural activation patterns underlying neural signals from similar tasks.
(2) Certain tasks from different domains exhibit strong similarities, suggesting shared neural mechanisms across cognitive functions. For example, clinical anomaly detection tasks (TUAB, TUEV) and emotion recognition tasks (SEED, SEED-IV) show similar spatial patterns,  due to overlapping activity in the bilateral temporal regions~\cite{zheng2015seed, SeizureChannel}, particularly around electrodes T7 and T8.
Similarly, sleep staging and cognitive workload tasks demonstrate over 75\% similarity in spatial query usage, due to converging neural activity in central and frontal     regions~\cite{moctezuma2024sleepchannel, cognitiveload_channel}. 
These findings reveal underlying neural correlations between tasks, which is consistent with existing neuroscientific research, implying the potential of shared representations for multi-task learning.

\begin{figure}[htbp]
  \centering
    \vspace{-2mm}
    \includegraphics[width=1\textwidth]{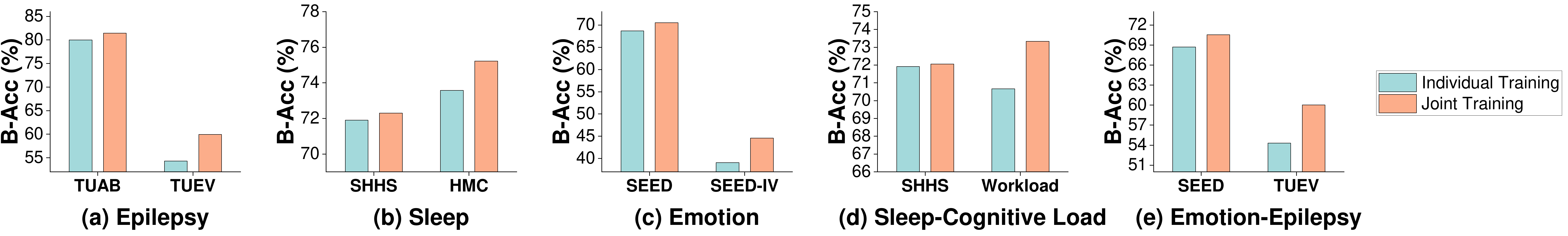}
  \vspace{-4mm}
  \caption{Comparison of individual and joint training balanced accuracy across multiple tasks.}
  \label{fig:joint-training}
  \vspace{-7mm}
\end{figure}

\textbf{Mutual Enhancement in Multi-task Training.} 
We further validate the mutual enhancement effect  in multi-task training by
selecting six pairs of datasets with high similarity scores.
As shown in Figure~\ref{fig:joint-training}, 
joint training consistently leads to better performance than training each task separately. For epilepsy-related tasks, TUEV gains 5.64\% with TUAB, SEED-IV improves 5.54\% with SEED, workload task benefits 2.66\% from sleep data, and TUEV improves 5.74\% when paired with SEED.
The results demonstrate that UniMind facilitates effective multi-task training, enhancing the performance across diverse decoding tasks by sharing neural representations. The improvement highlights the potential of multi-task foundation models in advancing generalized brain decoding.


\section{Conclusion}
In this paper, we present UniMind, a general-purpose EEG foundation model for multi-task brain decoding. It leverages a Neuro-Language Connector to align spatiotemporal EEG features with LLMs and a Task-aware Query Selection Module to adapt to diverse tasks. Experiments on ten datasets show UniMind outperforms prior models. We believe this work lays a solid foundation for future research on LLM–brain interaction and multi-task learning in the EEG domain. 

\textbf{Limitations and Future Work.} This work does not investigate the model’s generalization ability to completely unseen subjects and novel tasks, which may limit its ability to wider applications. Future efforts will focus on conducting cross-subject experiments in zero-shot or few-shot settings and integrating a broader range of EEG tasks to enhance the model’s robustness and generalizability. 


{\small
\bibliographystyle{unsrt}
\bibliography{main}
}

\newpage
\appendix
\section{Related Works}
\textbf{Task-Specific EEG Decoding Models.} 
Due to the variations in EEG signal formats across different datasets, numerous deep learning models have been proposed to tackle the aforementioned tasks within their respective domains. These models primarily focus on designing tailored feature extraction networks to accommodate EEG samples for specific tasks. Some works~\cite{jing2023SPaRCNet, 1d-cnnforeegsignals, dar2020cnnPhysiological} apply 1D convolutional neural networks (CNNs) directly to raw signals, while others~\cite{yang2022atd, kim2020ecg, kim2020ecg} first preprocess the data using short-time Fourier transform (STFT) and then adopt 2D CNNs on the resulting spectrograms. However, these methods offer a relatively narrow scope of perception, focusing mainly on local information. To address this, some methods~\cite{jing2020automated, ObstructiveSleepApnoea, yang2023Electroencephalogram, li2022FFCL} build on CNN-based encoders by incorporating Transformers, LSTMs, or recurrent neural networks (RNNs) to model temporal dynamics and capture global dependencies.
To further enhance EEG feature representation, some approaches employ multiple encoders for ensemble learning~\cite{li2022FFCL}, or leverage multi-level Transformers to encode spatial and temporal features both across and within channels\cite{song2021transformer, liu2022spatialtemporaltransformerseegemotion}. However, these task-specific models often suffer from limited generalization and poor transferability, restricting their applicability to broader EEG analysis tasks.

\textbf{EEG Foundation Model.} 
Because task-specific models can only handle a single task and lack cross-task learning capabilities, their broader applicability is limited. 
As a result, EEG foundation models have attracted increasing attention. These models aim to learn universal EEG representations by training on diverse datasets, enabling generalization across various downstream tasks. 
To address the challenges posed by heterogeneous EEG formats, Yang et al.~\cite{yang2023biot} proposed BIOT, which tokenizes EEG channels into fixed-length segments and incorporates channel and relative position embeddings to preserve spatio-temporal features.
Inspired by masked language modeling in LLMs, Jiang et al.~\cite{jiang2024Labram} extended this idea with LaBraM, introducing a neural tokenizer and pre-training it via masked neural code prediction on large-scale EEG data, achieving SOTA results across multiple tasks. 
While these models effectively address many challenges, they still require separate fine-tuning for each downstream task. NeuroLM~\cite{jiang2025neurolm} was introduced as the first multi-task foundation model for EEG, leveraging the capabilities of LLMs to support multi-task learning and inference. However, its performance is limited by the modality gap between EEG and language, as well as interference among tasks. In contrast, UniMind bridges this gap through a spatio-temporal cross-modality alignment strategy and reduces task interference via task-aware query selection. This design allows UniMind to achieve stronger performance across diverse tasks while supporting unified multi-task decoding within a single model.

\textbf{Multimodal Large Language Models.}
Building on the success of Large Language Models (LLMs), the Multimodal Large Language Models (MLLMs) are developed to enhance cross-modal understanding by combining visual, auditory, and textual data. Some apporaches~\cite{alayrac2022flamingo, awadalla2023OpenFlamingo, li2023MIMIC-IT, zhang2024LLaMA-Adapter} enhance LLMs by incorporating components like gated cross-attention layers or adapter layers to handle multimodal inputs. Others~\cite{dai2023InstructBLIP, liu2023LLaVA, zhu2023MiniGPT-4} use projection layers or Q-Formers to map visual encoder outputs into LLM input space. Video LLMs~\cite{li2024VideoChat, zhang2023Video-LLaMA, maaz2024Video-ChatGPT, luo2023Valley, ji2024wavchat} extend MLLMs for video tasks, primarily using projection layers or Q-Formers to process visual tokens. Besides, recent MLLMs have successfully integrated modalities such as audio~\cite{cheng2024VideoLLaMA2, su2023pandagpt, wu2024NextGPT}, applying them to EEG signals remains particularly challenging. While both audio and EEG exhibit continuous temporal patterns, EEG is inherently more complex due to its non-stationary characteristics across both temporal and spatial dimensions. It captures fine-grained neural activity that varies over time and across different brain regions, introducing significant variability and noise.
Unlike text, which is composed of discrete and structured tokens, EEG signals are fluid, high-dimensional, and constantly changing. 
This makes it difficult for LLMs to directly process and align with EEG representations, posing a major challenge for effective cross-modal integration.
Building on these insights, UniMind proposes the Neuro-Language Connector, a crucial module that bridges the substantial modality gap between continuous, complex EEG signals and the discrete token representations of LLMs. By encoding spatiotemporal neural patterns into interpretable features, it enables more effective cross-modal alignment, paving the way for improved integration of EEG within multimodal language models.

\section{Dataset Details}
\textbf{Dataset information.}
To comprehensively evaluate \textit{UniMind}, we utilize ten publicly available EEG datasets covering five representative task domains.
(1) For \textbf{sleep stage classification}, we employ HMC~\cite{alvarez2021hmc}, SleepEDF~\cite{kemp2000sleepedf}, and SHHS~\cite{quan1997shhs}. These datasets provide long-term polysomnographic EEG recordings annotated by experts, with each EEG segment labeled into one of five standard sleep stages: Wake, NREM-1, NREM-2, NREM-3, and REM.
(2) For \textbf{emotion recognition}, we adopt SEED~\cite{zheng2015seed} and SEED-IV~\cite{zheng2019seediv}, both containing multi-session EEG data collected while subjects viewed emotional video stimuli. SEED includes three emotion categories (positive, negative, neutral), whereas SEED-IV expands this to four categories: neutral, sad, fear, and happy.
(3) For \textbf{Clinical Abnormaly Detection}, we use TUAB~\cite{harati2015tuab}, TUEV~\cite{harati2015tuab}, and TUSL~\cite{von2017tusl}. TUAB focuses on detecting abnormal EEG activity, TUEV classifies six types of clinical events such as epileptiform discharges and eye movements, and TUSL categorizes seizure, slowing, and complex background activity.
(4) For \textbf{cognitive workload classification}, we utilize the Workload dataset~\cite{zyma2019electroencephalograms}, which contains EEG recordings collected during mental arithmetic tasks designed to distinguish between high and low cognitive load.
(5) Finally, for \textbf{motor imagery classification}, we incorporate SHU~\cite{ma2022shu}, a large-scale dataset featuring EEG signals corresponding to imagined left- and right-hand movements, enabling binary decoding of motor intentions in brain–computer interface applications.

\begin{table}[ht]
\vspace{-4mm}
\centering
\caption{Overview of Downstream EEG datasets and tasks used for evaluation.}
\renewcommand{\arraystretch}{0.85}
\resizebox{\textwidth}{!}{%
\begin{tabular}{l l c c c c c c}
\toprule
\textbf{Task} & \textbf{Dataset} & \textbf{Rate (Hz)} & \textbf{\#Channels} & \textbf{Duration} & \textbf{\#Samples} & \textbf{\#Subjects} & \textbf{Label} \\
\midrule
\multirow{1}{*}{Motor Imagery}   
& SHU      \cite{ma2022shu}                                            & 250   & 32  & 4s  & 11,988  & 25   & Binary-class \\
\midrule
\multirow{2}{*}{Emotion Recognition} & SEED       \cite{zheng2015seed}  & 1000   & 62  & 1s & 144,851  & 15  & 3-class \\
                                     & SEED-IV       \cite{zheng2019seediv}  & 1000  & 62  & 1s  & 143,610 & 16   & 5-class \\
\midrule
\multirow{3}{*}{Clinical Abnormaly Detection}      
& TUAB~\cite{harati2015tuab} & 250/256/512 & 23 & 10s & 409,083 & 2,383 & Binary-class \\
& TUEV~\cite{harati2015tuab} & 250 & 23 & 5s & 112,237 & 370  & 6-class \\
& TUSL~\cite{von2017tusl} & 250 & 23 & 10s & 245 & 28 & 3-class \\
\midrule
\multirow{2}{*}{Sleep Classification}       & SHHS~\cite{quan1997shhs} & 125 & 1 & 30s & 324,854  & 329  & 5-class \\
& SleepEDF~\cite{kemp2000sleepedf} & 100 & 1 & 30s & 414,961  & 78  & 5-class \\
& HMC~\cite{alvarez2021hmc}  & 256 & 4   & 30s & 137,243 & 151  & 4-class \\
\midrule
Workload Classification & Workload~\cite{zyma2019electroencephalograms} & 500 & 19 & 4s  & 1080 & 36 & Binary-class \\
\bottomrule
\end{tabular}%
}
\label{tab:dataset_summary}
\end{table}

\textbf{Data Splits.} Each dataset is divided into training, validation, and test sets according to task-specific strategies. (1) \textbf{TUAB and TUEV}: we follow the official training and test split, and further divide the training subjects into 80\% for training and 20\% for validation. (2) \textbf{SEED and SEED-IV}: trials are split chronologically. SEED contains 15 trials, divided into 9 for training, 3 for validation, and 3 for testing; SEED-IV contains 24 trials, divided into 16, 4, and 4 respectively. All sessions within each split are merged. (3) \textbf{HMC}: subjects 1 to 100 are assigned to the training set, 101 to 126 to the validation set, and 127 to 154 to the test set. (4) \textbf{Workload}: subjects 0 to 25 are used for training, 26 to 30 for validation, and 31 to 35 for testing. (5) \textbf{TUSL}: the dataset is randomly divided into 60\% for training, 20\% for validation, and 20\% for testing. (6) \textbf{SleepEDF and SHHS}: subjects are randomly assigned to training, validation, and test sets in an 8:1:1 ratio. (7) \textbf{SHU}: sessions are split into training, validation, and test sets in a 3:1:1 ratio.

\textbf{Data Preprocessing.} 
We develop a standardized data pre-processing pipeline comprising band-pass filtering, notch filtering, resampling, and normalization, designed to be universally applicable across various BCI tasks and decoding models. Following established practice, we apply a 0.1–75 Hz band-pass filter to retain task-relevant frequency components while suppressing low-frequency drifts and high-frequency noise.To remove power-line interference, we first perform a Fast Fourier Transform (FFT) on the raw EEG data to identify the specific interference frequency, and then apply a notch filter at 50 Hz or 60 Hz based on the local power-line frequency. Finally, all EEG signals are resampled to 200 Hz. To mitigate the effects of low-amplitude signals and enhance numerical stability during optimization, we apply z-score normalization to most datasets. Specifically, for the SHU dataset, we instead use 95\% normalization due to its signal distribution characteristics.

\textbf{Metrics.}
Due to the inherent class imbalance in EEG datasets, we evaluate model performance using three primary metrics. \textbf{(1) Balanced Accuracy} is defined as the average recall obtained on each class. By equally weighting the recall of all classes, it effectively mitigates the bias caused by imbalanced class distributions, ensuring that minority classes are not overlooked during evaluation.
\textbf{(2) Cohen’s Kappa} measures the level of agreement between the predicted labels and the true labels, while correcting for the agreement that could occur by random chance. This statistic provides a more robust assessment of classification performance than simple accuracy, especially in datasets where class distribution is uneven.
\textbf{(3) Weighted F1 Score} is the harmonic mean of precision and recall calculated for each class individually and then averaged by weighting each class's score according to its support (i.e., the number of true instances). This metric balances false positives and false negatives while accounting for class imbalance, providing a comprehensive measure of overall model effectiveness.

Since our model produces discrete class labels directly from LLM outputs without associated probability scores, metrics that require confidence values such as \textbf{AUROC} (Area Under the Receiver Operating Characteristic Curve) and \textbf{AUC-PR} (Area Under the Precision-Recall Curve) are not applicable. Therefore, \textbf{Weighted F1} and \textbf{Cohen’s Kappa} are prioritized for both binary and multi-class classification tasks to ensure reliable evaluation under these constraints.

\section{Training Details}
\textbf{Experimental Setup.} All experiments are conducted on a single machine equipped with eight NVIDIA A800 GPUs, each with 80GB of memory. We adopt a batch-wise alternating training strategy, where each mini-batch is drawn from a single dataset. The software environment includes Python 3.9 and PyTorch 2.0.1 with CUDA 11.8 support. We train our model using 8-GPU data parallelism with a total effective batch size of 128. Training is performed for 10 epochs using the AdamW optimizer with a cosine learning rate schedule, a base learning rate of 4e-5, and a warmup ratio of 3\%. 
Please refer to Table~\ref{tab:training_config} for additional training details.
\begin{table}[htbp]
\vspace{-4mm}
\centering
\caption{Key training and model configuration parameters}
\begin{tabular}{ll}
\toprule
\textbf{Parameter} & \textbf{Value} \\
\midrule
\quad Training Epochs & 10 \\
\quad Batch Size & 128 \\
\quad Learning Rate & $4 \times 10^{-5}$ \\
\quad Weight Decay & 0.01 \\
\quad Warmup Ratio & 0.03 \\
\quad LR Scheduler & Cosine decay \\
\midrule
\quad Max Sequence Length & 4096 \\
\quad Dataloader Workers & 4 \\
\quad BFloat16 Precision & True \\
\quad Use Thumbnail & True \\
\midrule
\quad EEG Hidden Size & 1152 \\
\quad Number of EEG Encoder Layers & 12 \\
\quad Number of EEG Attention Heads & 10 \\
\quad LLM Hidden Size & 4096 \\
\quad Number of LLM Layers & 32 \\
\quad Number of LLM Attention Heads & 32 \\
\quad RoPE Scaling & Dynamic, factor = 2.0 \\
\bottomrule
\vspace{-2mm}
\end{tabular}
\label{tab:training_config}
\end{table}

\textbf{Computational Requirements.} Our language models adopt three different sizes, corresponding to InternLM-0.5B, InternLM-1.8B, and InternLM-7B as shown in Table~\ref{tab:internlm_resource}. During training, we freeze the LLM parameters and only fine-tune other components, with a total of 87,592,512 trainable parameters. The table reports the training time and maximum GPU memory consumption on 8 NVIDIA A800 GPUs.
\begin{table}[htbp]
\vspace{-2mm}
\centering
\caption{Training resource comparison of UniMind}
\label{tab:internlm_resource}
\begin{tabular}{lcc}
\toprule
\textbf{Language Model Size} & \textbf{Time (8-GPU, hours)} & \textbf{Max GPU Memory (GB)} \\
\midrule
InternLM-0.5B & 7.34 & 31.2 \\
InternLM-1.8B & 8.37 & 39.4 \\
InternLM-7B   & 21.78 & 61.7 \\
\bottomrule
\vspace{-8mm}
\end{tabular}
\end{table}

\section{More Results.}
We present the complete results for Balanced Accuracy, Cohen’s Kappa, and Weighted F1 across all ten datasets. By conducting multiple experiments with different random seeds, we calculate the mean values along with their standard deviations. The outcomes on these three metrics are generally consistent with the analysis provided in the main text. Additionally, UniMind demonstrates relatively smaller standard deviations compared to other models, especially on the smaller datasets such as TUSL and Workload, which indicates more stable and reliable performance.

\begin{table}[ht]
\centering
\caption{Performance comparison on SEED~\cite{zheng2015seed} and HMC~\cite{alvarez2021hmc} datasets. “MT” denotes multi-task learning usage. \underline{Underlining} shows best multi-task results; \textbf{bold} shows best overall.}
\label{tab:seed_hmc_horizontal}
\renewcommand{\arraystretch}{0.95}
\scriptsize
\begin{tabular}{l|c|ccc|ccc}
\toprule
\textbf{Model} & \textbf{MT} & \multicolumn{3}{c|}{\textbf{SEED}} & \multicolumn{3}{c}{\textbf{HMC}} \\ &
              & B-Acc & Kappa & F1-W & B-Acc & Kappa & F1-W \\
\midrule
SPaRCNet  & \ding{53} & 55.96 ($\pm$2.44) & 34.64 ($\pm$3.72) & 55.85 ($\pm$2.97) & 47.56 ($\pm$11.09) & 31.47 ($\pm$13.15) & 41.08 ($\pm$13.10) \\
ContraWR & \ding{53} & 61.06 ($\pm$0.78) & 42.20 ($\pm$1.29) & 61.37 ($\pm$0.85) & 42.42 ($\pm$5.41) & 23.40 ($\pm$5.54) & 29.87 ($\pm$2.88) \\
CNN-Trans. & \ding{53} & 61.61 ($\pm$3.84) & 42.62 ($\pm$6.01) & 61.50 ($\pm$4.63) & 65.73 ($\pm$1.41) & 59.61 ($\pm$1.05) & 68.96 ($\pm$0.65) \\
FFCL &\ding{53}           & 58.08 ($\pm$3.22) & 37.32 ($\pm$4.62) & 57.43 ($\pm$4.02) & 44.27 ($\pm$7.02) & 25.42 ($\pm$6.54) & 29.02 ($\pm$4.85) \\
ST-Trans. & \ding{53} & 54.79 ($\pm$0.91) & 32.61 ($\pm$1.69) & 55.05 ($\pm$0.91) & 25.59 ($\pm$1.41) & 5.03 ($\pm$1.83)  & 14.28 ($\pm$1.22) \\
BIOT & \ding{53} & 70.97 ($\pm$0.24) & 56.82 ($\pm$0.51) & 71.34 ($\pm$0.27) & 68.62 ($\pm$0.41) & 62.95 ($\pm$1.13) & 70.91 ($\pm$1.47) \\
LaBraM & \ding{53} & \textbf{73.18} ($\pm$0.19) & \textbf{59.94} ($\pm$0.31) & \textbf{73.54} ($\pm$0.21) & 72.86 ($\pm$1.01) & 68.12 ($\pm$0.73) & 75.54 ($\pm$0.24) \\
NeuroLM & \ding{51} & 60.34 ($\pm$0.10) & 40.82 ($\pm$0.36) & 60.63 ($\pm$0.30) & 67.37 ($\pm$0.50) & 61.88 ($\pm$0.57) & 71.26 ($\pm$0.34) \\
UniMind & \ding{51} & \underline{70.55} ($\pm$0.61) & \underline{56.28} ($\pm$0.73) & \underline{70.98} ($\pm$0.96) & \underline{\textbf{75.27}} ($\pm$0.56) & \underline{\textbf{70.58}} ($\pm$0.48) & \underline{\textbf{77.40}} ($\pm$0.73) \\
\bottomrule
\end{tabular}
\end{table}

\begin{table}[ht]
\centering
\caption{Performance comparison on TUAB~\cite{harati2015tuab} and TUEV~\cite{harati2015tuab} datasets. “MT” denotes multi-task learning usage. \underline{Underlining} shows best multi-task results; \textbf{bold} shows best overall.}
\label{tab:tuab_tuev_horizontal}
\renewcommand{\arraystretch}{0.95}
\scriptsize
\begin{tabular}{l|c|ccc|ccc}
\toprule
\textbf{Model} & \textbf{MT} & \multicolumn{3}{c|}{\textbf{TUAB}} & \multicolumn{3}{c}{\textbf{TUEV}} \\ &
              & B-Acc & Kappa & F1-W & B-Acc & Kappa & F1-W \\
\midrule
SPaRCNet  & \ding{53} & 78.69 ($\pm$0.47) & 50.67 ($\pm$0.29) & 75.13 ($\pm$0.80) & 41.61 ($\pm$2.62) & 42.33 ($\pm$1.81) & 70.24 ($\pm$1.04) \\
ContraWR & \ding{53} & 80.17 ($\pm$0.58) & 61.02 ($\pm$0.38) & 80.65 ($\pm$0.42) & 43.84 ($\pm$3.49) & 39.12 ($\pm$2.37) & 68.93 ($\pm$1.36) \\
CNN-Trans. & \ding{53} & 79.53 ($\pm$0.94) & 57.33 ($\pm$0.63) & 78.76 ($\pm$0.31) & 40.87 ($\pm$1.61) & 38.15 ($\pm$1.34) & 68.54 ($\pm$2.93) \\
FFCL &\ding{53} & 78.19 ($\pm$0.22) & 55.81 ($\pm$0.84) & 77.83 ($\pm$0.75) & 39.79 ($\pm$1.04) & 37.32 ($\pm$1.88) & 67.83 ($\pm$1.20) \\
ST-Trans. & \ding{53} & 79.66 ($\pm$0.31) & 61.69 ($\pm$0.45) & 80.90 ($\pm$0.93) & 39.84 ($\pm$2.28) & 37.65 ($\pm$3.06)  & 68.23 ($\pm$1.90) \\
BIOT & \ding{53} & 79.59 ($\pm$0.89) & 59.42 ($\pm$0.13) & 78.82 ($\pm$0.90) & 52.81 ($\pm$2.25) & 52.73 ($\pm$2.49) & 74.92 ($\pm$0.82) \\
LaBraM & \ding{53} & 81.40 ($\pm$0.20) & 62.52 ($\pm$0.12) & 81.47 ($\pm$0.57) & \textbf{64.09} ($\pm$0.65) & \textbf{66.37} ($\pm$0.93) & \textbf{83.12} ($\pm$0.52) \\
NeuroLM & \ding{51} & 79.69 ($\pm$0.53) & 54.61 ($\pm$0.98) & 78.93 ($\pm$0.91) & 46.79 ($\pm$3.56) & 45.70 ($\pm$4.98) & 73.59 ($\pm$2.19) \\
UniMind & \ding{51} & \underline{\textbf{81.76}} ($\pm$0.33) & \underline{\textbf{63.80}} ($\pm$0.77) & \underline{\textbf{82.03}} ($\pm$0.60) & \underline{63.19} ($\pm$1.93) & \underline{56.29} ($\pm$1.11) & \underline{78.04} ($\pm$0.84) \\
\bottomrule
\end{tabular}
\end{table}

\begin{table}[ht]
\centering
\caption{Performance comparison on TUSL~\cite{zheng2019seediv} and Workload~\cite{kemp2000sleepedf} datasets. “MT” denotes multi-task learning usage. \underline{Underlining} shows best multi-task results; \textbf{bold} shows best overall.}
\label{tab:tusl_workload_horizontal}
\renewcommand{\arraystretch}{0.95}
\scriptsize
\begin{tabular}{l|c|ccc|ccc}
\toprule
\textbf{Model} & \textbf{MT} & \multicolumn{3}{c|}{\textbf{TUSL}} & \multicolumn{3}{c}{\textbf{Workload}} \\
              &             & B-Acc & Kappa & F1-W & B-Acc & Kappa & F1-W \\
\midrule
SPaRCNet  & \ding{53} & 41.85 ($\pm$4.52) & 13.99 ($\pm$7.99) & 35.00 ($\pm$9.68) & 59.77 ($\pm$3.67) & 10.67 ($\pm$2.42) & 54.60 ($\pm$4.11) \\
ContraWR & \ding{53} & 58.57 ($\pm$6.62) & 35.67 ($\pm$9.68) & 54.58 ($\pm$7.98) & 69.66 ($\pm$4.33) & 31.08 ($\pm$3.81) & 69.33 ($\pm$3.78) \\
CNN-Trans. & \ding{53} & 35.75 ($\pm$3.51) & 3.06 ($\pm$4.79) & 22.35 ($\pm$2.51) & 57.93 ($\pm$5.14) & 8.37 ($\pm$4.35) & 58.63 ($\pm$4.62) \\
FFCL & \ding{53} & 39.19 ($\pm$6.88) & 6.28 ($\pm$8.88) & 21.20 ($\pm$7.86) & 70.69 ($\pm$2.46) & 41.36 ($\pm$3.12) & 72.54 ($\pm$2.93) \\
ST-Trans. & \ding{53} & 40.00 ($\pm$3.29) & 8.60 ($\pm$4.49) & 37.93 ($\pm$4.59) & 61.03 ($\pm$1.87) & 12.48 ($\pm$2.33) & 60.67 ($\pm$1.52) \\
BIOT & \ding{53} & 57.58 ($\pm$3.03) & 20.12 ($\pm$2.12) & 23.94 ($\pm$0.40) & 66.55 ($\pm$2.74) & 30.67 ($\pm$3.30) & 51.66 ($\pm$2.61) \\
LaBraM & \ding{53} & 76.25 ($\pm$2.31) & \textbf{64.07 ($\pm$3.04)} & \textbf{76.14 ($\pm$2.12)} & 66.09 ($\pm$1.94) & 29.81 ($\pm$1.35) & 64.72 ($\pm$2.43) \\
NeuroLM & \ding{51} & 68.45 ($\pm$3.04) & 51.94 ($\pm$4.61) & 68.39 ($\pm$2.97) & 63.45 ($\pm$2.25) & 23.19 ($\pm$1.74) & 66.57 ($\pm$1.92) \\
UniMind & \ding{51} & \underline{\textbf{78.95 ($\pm$1.34)}} & \underline{61.53 ($\pm$1.58)} & \underline{75.40 ($\pm$1.77)} & \underline{\textbf{78.67 ($\pm$1.43)}} & \underline{\textbf{57.33 ($\pm$1.12)}} & \underline{\textbf{78.65 ($\pm$1.67)}} \\
\bottomrule
\end{tabular}
\end{table}

\begin{table}[ht]
\centering
\caption{Performance comparison on SEED-IV~\cite{zheng2019seediv} and SleepEDF~\cite{kemp2000sleepedf} datasets. “MT” denotes multi-task learning usage. \underline{Underlining} shows best multi-task results; \textbf{bold} shows best overall.}
\label{tab:seediv_sleepedf_horizontal}
\renewcommand{\arraystretch}{0.95}
\scriptsize
\begin{tabular}{l|c|ccc|ccc}
\toprule
\textbf{Model} & \textbf{MT} & \multicolumn{3}{c|}{\textbf{SEED-IV}} & \multicolumn{3}{c}{\textbf{SleepEDF}} \\
              &             & B-Acc & Kappa & F1-W & B-Acc & Kappa & F1-W \\
\midrule
SPaRCNet  & \ding{53} & 29.88 ($\pm$1.37) & 6.45 ($\pm$0.92) & 32.05 ($\pm$1.48) & 60.16 ($\pm$1.21) & 65.31 ($\pm$1.36) & 58.61 ($\pm$1.44) \\
ContraWR & \ding{53} & 38.38 ($\pm$1.61) & 14.83 ($\pm$1.26) & 40.21 ($\pm$1.43) & 67.05 ($\pm$1.67) & 70.46 ($\pm$1.52) & 66.92 ($\pm$1.58) \\
CNN-Trans. & \ding{53} & 35.21 ($\pm$1.42) & 10.92 ($\pm$1.17) & 36.57 ($\pm$1.39) & 60.29 ($\pm$1.35) & 66.12 ($\pm$1.21) & 58.96 ($\pm$1.30) \\
FFCL & \ding{53} & 37.81 ($\pm$1.55) & 15.35 ($\pm$1.44) & 39.76 ($\pm$1.61) & 65.66 ($\pm$1.33) & 70.13 ($\pm$1.27) & 64.79 ($\pm$1.45) \\
ST-Trans. & \ding{53} & 36.93 ($\pm$1.46) & 15.72 ($\pm$1.30) & 36.95 ($\pm$1.50) & 55.17 ($\pm$1.49) & 67.32 ($\pm$1.38) & 53.18 ($\pm$1.42) \\
BIOT & \ding{53} & 36.19 ($\pm$1.38) & 23.21 ($\pm$1.47) & 42.76 ($\pm$1.52) & 64.95 ($\pm$0.26) & 71.32 ($\pm$0.32) & 60.91 ($\pm$0.40) \\
LaBraM & \ding{53} & \textbf{47.63 ($\pm$0.73)} & \textbf{28.77 ($\pm$0.60)} & \textbf{49.14 ($\pm$0.82)} & 68.96 ($\pm$0.75) & 75.49 ($\pm$0.68) & 87.30 ($\pm$0.71) \\
NeuroLM & \ding{51} & 32.30 ($\pm$1.69) & 9.46 ($\pm$1.53) & 34.65 ($\pm$1.66) & 56.40 ($\pm$0.60) & 68.76 ($\pm$0.59) & 54.02 ($\pm$0.63) \\
UniMind & \ding{51} & \underline{45.56 ($\pm$0.71)} & \underline{24.06 ($\pm$0.58)} & \underline{43.58 ($\pm$0.77)} & \underline{\textbf{72.98 ($\pm$0.66)}} & \underline{\textbf{76.95 ($\pm$0.63)}} & \underline{\textbf{88.23 ($\pm$0.69)}} \\
\bottomrule
\end{tabular}
\end{table}

\begin{table}[ht]
\centering
\caption{Performance comparison on SHHS~\cite{quan1997shhs} and SHU~\cite{ma2022shu} datasets. “MT” denotes multi-task learning usage. \underline{Underlining} shows best multi-task results; \textbf{bold} shows best overall.
}
\label{tab:shhs_shu_horizontal}
\renewcommand{\arraystretch}{0.95}
\scriptsize
\begin{tabular}{l|c|ccc|ccc}
\toprule
\textbf{Model} & \textbf{MT} & \multicolumn{3}{c|}{\textbf{SHHS}} & \multicolumn{3}{c}{\textbf{SHU}} \\
              &             & B-Acc & Kappa & F1-W & B-Acc & Kappa & F1-W \\
\midrule
SPaRCNet      & \ding{53} & 63.93 ($\pm$0.97) & 63.47 ($\pm$0.91) & 61.82 ($\pm$0.95) & 62.15 ($\pm$0.78) & 25.32 ($\pm$0.81) & 62.15 ($\pm$0.74) \\
ContraWR     & \ding{53} & 67.01 ($\pm$1.02) & 69.44 ($\pm$1.09) & 56.80 ($\pm$0.96) & 62.13 ($\pm$0.83) & 24.15 ($\pm$0.88) & 57.51 ($\pm$0.86) \\
CNN-Trans.    & \ding{53} & 64.63 ($\pm$0.92) & 67.83 ($\pm$0.87) & 63.78 ($\pm$0.90) & 56.25 ($\pm$0.76) & 12.48 ($\pm$0.80) & 55.86 ($\pm$0.73) \\
FFCL          & \ding{53} & 67.59 ($\pm$0.89) & 69.48 ($\pm$0.93) & 67.07 ($\pm$0.90) & 62.82 ($\pm$0.74) & 25.67 ($\pm$0.78) & 62.78 ($\pm$0.76) \\
ST-Trans.     & \ding{53} & 64.63 ($\pm$0.95) & 67.18 ($\pm$0.91) & 63.30 ($\pm$0.94) & 63.39 ($\pm$0.70) & 26.76 ($\pm$0.73) & 63.25 ($\pm$0.75) \\
BIOT          & \ding{53} & 72.22 ($\pm$0.73) & \textbf{78.35 ($\pm$0.78)} & 83.96 ($\pm$0.76) & 59.16 ($\pm$0.61) & 16.68 ($\pm$0.66) & 55.51 ($\pm$0.64) \\
LaBraM        & \ding{53} & 71.69 ($\pm$0.68) & 77.07 ($\pm$0.72) & 82.90 ($\pm$0.69) & \textbf{67.90 ($\pm$0.64)} & \textbf{32.85 ($\pm$0.66)} & \textbf{67.84 ($\pm$0.63)} \\
NeuroLM      & \ding{51} & 59.15 ($\pm$1.23) & 61.68 ($\pm$1.15) & 63.54 ($\pm$1.10) & 59.36 ($\pm$0.78) & 16.91 ($\pm$0.75) & 56.03 ($\pm$0.73) \\
UniMind     & \ding{51} & \underline{\textbf{74.00 ($\pm$0.62)}} & \underline{76.63 ($\pm$0.66)} & \underline{\textbf{84.20 ($\pm$0.69)}} & \underline{65.77 ($\pm$0.60)} & \underline{31.54 ($\pm$0.63)} & \underline{65.73 ($\pm$0.59)} \\
\bottomrule
\end{tabular}
\end{table}

\section{Instruction Construction}

We adopt instruction tuning to train UniMind. To enhance instruction diversity, we design ten distinct prompts for each dataset, with each sample paired with one randomly selected task-specific instruction. Each sample contains a unique identifier, a label, the file path to the corresponding EEG signal, and a simulated human-AI dialogue. The conversation begins with the human posing a question about the category represented by the EEG signal, using one of ten predefined templates selected at random. The GPT model then generates a response based on the associated label. This dataset is organized as natural language dialogues, making it well-suited for instruction tuning and improving the model’s ability to interpret the semantics of EEG signals.


\begin{tcolorbox}[colback=gray!5!white,colframe=gray!95!black,title=Instruction Templates for the SHU Dataset]
\begin{enumerate}
  \item This segment of EEG signal can reflect the subject's behavioral actions. Please determine the type of action based on the provided EEG signal? [Left hand, Right hand]
  \item The given EEG signal is indicative of the subject's movements. Can you identify the action type from the EEG data? [Left hand, Right hand]
  \item Analyze this EEG signal to discern the subject's physical actions. What is the action type shown? [Left hand, Right hand]
  \item ... (similar instructions)
\end{enumerate}
\end{tcolorbox}

\begin{tcolorbox}[colback=gray!5!white,colframe=gray!95!black,title=Instruction Templates for the SEED Dataset]
\begin{enumerate}
  \item  Given this EEG signal, which emotion does it reflect? [positive, negative, or neutral]
  \item  Based on this EEG signal, please identify the emotion it represents. [positive, negative, or neutral]
  \item  From this EEG signal, can you determine which emotion it corresponds to? [positive, negative, or neutral]
  \item ... (similar instructions)
\end{enumerate}
\end{tcolorbox}

\begin{tcolorbox}[colback=gray!5!white,colframe=gray!95!black,title=Instruction Templates for the SEED-IV Dataset]
\begin{enumerate}
  \item Given this EEG signal, which emotion does it reflect? [neutral, sad, fear, happy]
 \item Based on this EEG signal, please identify the emotion it represents. [neutral, sad, fear, happy]
 \item From this EEG signal, can you determine which emotion it corresponds to? [neutral, sad, fear, happy]
 \item ... (similar instructions)
\end{enumerate}
\end{tcolorbox}

\begin{tcolorbox}[colback=gray!5!white,colframe=gray!95!black,title=Instruction Templates for the TUAB Dataset]
\begin{enumerate}
  \item This EEG signal may indicate abnormal conditions. Based on this signal, determine if there is an abnormality. Choose one: [Normal, Abnormal]
  \item Analyze this EEG signal to assess whether it reflects an abnormal condition. Please select one: [Normal, Abnormal]
  \item This EEG signal could suggest abnormal brain activity. Determine if the signal is normal or abnormal: [Normal, Abnormal]
  \item ... (similar instructions)
\end{enumerate}
\end{tcolorbox}

\begin{tcolorbox}[colback=gray!5!white,colframe=gray!95!black,title=Instruction Templates for the TUEV Dataset]
\begin{enumerate}
  \item This EEG signal reflects epileptic events. Please determine the epileptic state based on this signal.
  \item Analyze this EEG signal to classify the epileptic state. 
  \item This EEG signal may indicate epileptic activity. Based on the signal, identify the epileptic state. 
  \item ... (similar instructions)
\end{enumerate}
\end{tcolorbox}

\begin{tcolorbox}[colback=gray!5!white,colframe=gray!95!black,title=Instruction Templates for the TUSL Dataset]
\begin{enumerate}
  \item This EEG signal reflects a slow event. Based on this signal, please determine the state. Choose one: [bckg, seiz, slow]
  \item Analyze this EEG signal to classify the state it indicates. Select one: [bckg, seiz, slow]
  \item This EEG signal may suggest a slow event. Determine the corresponding state from the options: [bckg, seiz, slow]
  \item ... (similar instructions)
\end{enumerate}
\end{tcolorbox}

\begin{tcolorbox}[colback=gray!5!white,colframe=gray!95!black,title={Instruction Templates for the SHHS , SleepEDF and HMC Dataset}]
\begin{enumerate}
  \item The EEG signal provides insights into sleep stages. Which sleep phase does it most likely correspond to? Choose one: [Sleep stage W, Sleep stage N1, Sleep stage N2, Sleep stage N3, Sleep stage R]
  \item Sleep phases can be inferred from EEG signals. Given the signal, which phase is it most likely indicating? Pick one: [Sleep stage W, Sleep stage N1, Sleep stage N2, Sleep stage N3, Sleep stage R]
  \item This EEG signal reflects brain activity during sleep. Which sleep stage does it most likely represent? Select one: [Sleep stage W, Sleep stage N1, Sleep stage N2, Sleep stage N3, Sleep stage R]
  \item ... (similar instructions)
\end{enumerate}
\end{tcolorbox}

\begin{tcolorbox}[colback=gray!5!white,colframe=gray!95!black,title=Instruction Templates for the Workload Dataset]
\begin{enumerate}
  \item This is an EEG signal. Is this brainwave showing high workload or low workload? [high, low]
  \item Here's an EEG signal. Does it represent a high workload or a low workload on the brain? [high, low]
  \item This EEG signal is given. Is the workload indicated here high or low? [high, low]
  \item ... (similar instructions)
\end{enumerate}
\end{tcolorbox}

\section{Discussion}
In this work, UniMind establishes a promising approach for unified multi-task EEG decoding by leveraging a Neuro-Language Connector to align complex spatiotemporal EEG features with large language models, and a Task-aware Query Selection Module to dynamically adapt to heterogeneous decoding tasks. Our extensive experiments across ten datasets demonstrate that UniMind consistently outperforms previous multi-task models and achieves comparable or superior results to single-task approaches, highlighting the effectiveness of integrating LLMs in brain signal decoding fOr multi-task learning.

Despite these encouraging results, several limitations remain. First, the current evaluation focuses on datasets and tasks seen during training, without fully exploring the model’s ability to generalize to entirely unseen subjects or novel EEG tasks. 
Second, due to the limited amount of available data for visual and textual decoding tasks (fewer than 10k samples), these tasks were not included in the training stage. Unlike simpler tasks such as sleep stage classification, visual and textual decoding require richer semantic understanding and finer neural representation learning, which demands larger datasets for effective training.
Third, our work focuses exclusively on EEG data and does not consider other brain-computer interface modalities such as fMRI or MEG. Integrating these complementary neural signals may provide richer information but also introduces additional challenges that remain to be addressed in future work.


\textbf{Future Work.} Future research will focus on enhancing UniMind’s ability to generalize to unseen subjects and novel EEG tasks, including cross-subject zero-shot and few-shot learning scenarios. Furthermore, we plan to incorporate visual and textual decoding tasks into the training process. Compared to simpler classification tasks, these decoding tasks require capturing richer semantic information and more complex neural representations, which are crucial for advancing the model’s understanding of high-level cognitive processes. Integrating these tasks will significantly enhance UniMind’s semantic comprehension and extend its applicability to more sophisticated brain decoding scenarios. Finally, we aim to progressively extend the framework to include other neural modalities such as fMRI and MEG, building a more comprehensive and robust foundation model for multi-modal brain decoding.



Overall, UniMind provides a solid foundation and new insights for future exploration of large language models in multi-task brain decoding, marking an important step toward more generalizable and interpretable neuro-AI systems.

\appendix

\end{document}